\documentclass{aastex631}

\usepackage{amsmath}
\usepackage{multirow}
\usepackage{graphicx}
\usepackage{float}
\usepackage{color}

\shorttitle{The blazar OJ~287 jet from parsec- to kiloparsec-scales}
\shortauthors{M.S. Butuzova}

\begin{document}

\title{The blazar OJ~287 jet from parsec- to kiloparsec-scales}

\correspondingauthor{Marina Butuzova}
\email{mbutuzova@craocrimea.ru}

\author[0000-0001-7307-2193]{Marina S. Butuzova}
\affiliation{Crimean Astrophysical Observatory, Nauchny 298409, Crimea, Russia}

\begin{abstract}
  The curved shape of the kiloparsec-scale jet of the blazar OJ~287 is analyzed in the framework of the precession of the central engine, on the existence on which a large number of studies over the past decades are based.
  The data necessary for the analysis on the kiloparsec-scale jet velocity and angle with the line of sight are obtained based on two competing assumptions about the X-ray emission mechanism of the OJ~287 jet.
  Namely, there were both the inverse Compton scattering of the microwave background under the assumption of relativistic kiloparsec-scale jet and the inverse Compton scattering of the central source radiation.
  For the latter one, we showed that the expected flux from the kiloparsec-scale jet in the gamma range does not exceed the limit set for it according to \textit{Fermi}-LAT data. 
  We found that only the period of the kiloparsec-scale jet helix, estimated in the framework of the inverse Compton scattering of the central source radiation, agrees with the precession period of the central engine, determined from the modulation of the peak values of 12-year optical flares. 
\end{abstract}

\section{Introduction} \label{sec:intro}

The blazar OJ~287 ($z=0.306$) is remarkable that there are powerful repeated approximately every 12~years flares in the optical light curve spanning more than 100 years \citep[see, e.g.,][]{Sillanpaa88, Villforth10}, based on this \citet{Sillanpaa88} assumed that there is a binary supermassive black hole (BH) in the center of OJ~287.
The low-mass component, passing through the pericenter, causes a 12-year flare by the exert a tidal effect on the accretion disk of the primary BH.
Since the radiation, generated in an ultrarelativistic jet directed at a small angle to the line of sight, is significantly amplified in the observer's reference frame and the flux density of this radiation prevails over the flux density of the radiation, formed in other parts of the active galaxy, it is natural to expect that the jet is responsible for these flares.
A similar scenario was considered, for example, by \citet{VR98, Katz97, LehtoValtonen96}. 
\citet{VR98} assumed that there are two BH of the same mass, having an accretion disk and a jet.
These jets are curved and twisted. From the beginning, a small angle of the radiating region with the line of sight is reached in one jet, then in the second, that leads to the observed two-peak structure of 12-year flares. 
\citet{Katz97} considers a helical jet formed by the precession of the accretion disk of the primary BH caused by the orbital motion of the secondary BH.
In this framework, the second peak during the 12-year flare is assumed to be formed due to nutation.
Alternatively, to explain the two-peak structure of the 12-year flare, \citet{LehtoValtonen96} assume that the secondary BH passes through the accretion disk of the primary BH twice during the period.
At that, the primary BH has a mass of $1.8\cdot10^{10} M_\odot$, which is an order of magnitude higher than the estimates of the BH masses for blazars \citep[see, e.g.,][and references therein]{SbarratoGhM12,TitarchukSeifina17}.
Recent studies of the spectral energy distribution during a flare of 2015$-$2016 reveal indications of the contribution of the accretion disk radiation to the total OJ~287 emission \citep{Kushwaha18, Kushwaha20}.
The model of \cite{LehtoValtonen96} was supplemented with a precessing jet, which resulted in the agreement of the difference in the position angles (PA) of the jet \citep{ValtonenPihajoki13} observed with the VLBA at frequencies 15 and 43~GHz.
However, the prediction made by \citet{ValtonenPihajoki13} about the values of the position jet angles was not confirmed by further observations \citep{Cohen17, Agudo12, ButP20}.
On the other hand, \citet{Britzen18} explained the changes in PA of the parsec-scale (pc-) jet features by the precession of the jet with the nutational fluctuations superimposed on it.
In this case, the precession period of the jet PA ($\approx$22~years) is consistent with the long-term quasi-period of the radio flux variability ($\approx$25~years), which occurs due to periodic changes in the Doppler factor of the emitting region.
Taking into account that almost all of the radio emission observed on single-dish antennae is generated in the VLBI core \citep{Kovalev05}, which is part of the pc-scale jet, where the medium becomes optically transparent for radiation at a given frequency \citep{Pushkarev12}, the result of \citet{Britzen18} is self-consistent, but differs from the period in the optical range.
\citet{ButP20} explained this difference under the assumption that optical emission is formed closer to the true pc-scale jet base than radio emission and that the components of the helical jet move at a certain small angle to the radial direction.
At the same time, \citet{ButP20} shown that the jet with such geometric and kinematic parameters can forms due to the development of (magneto)hydrodynamic instabilities, for example, the Kelvin-Helmholtz instability \citep[see, e.g.,][]{Hardee82}.
Meanwhile, the true precession of the helical jet manifests itself in different values of the maximum flux and in the difference in the time interval between adjacent 12-year flares. This scenario gives the precession period of the OJ~287 central engine of 1200 years, which can be interpreted by the Lense-Thirring precession in the system of a single supermassive black hole \citep{ButP20}.

Evidence of the precession of the central engine in the active galactic nuclei may be present in kiloparsec-scale (kpc-) jets as their bends \citep[see, e.g.,][]{ApplSV96}. 
In the VLA observations at a frequency of 1.4~GHz, a kiloparsec-scale jet for OJ~287 is detected, which morphologically corresponds to the FR~I radio source, has a length of more than $25^{\prime\prime}$ and exhibits a bend of $15^{\prime\prime}$ \citep{PerlmanStocke94}.
With an increase in the observation frequency, the flux density from the kpc-scale jet decreases, and the jet is unavailable for observations at frequencies $\geq$15~GHz \citep{MarscherJ11}.
In the optical and infrared ranges, the kpc-scale jet is undetected \citep{Yanny97, MarscherJ11}. In the X-ray range, the jet extends up to a distance of $\approx$20$^{\prime\prime}$ from the core.
At the distance of $8^{\prime\prime}$, corresponding to the end of the relatively bright and near to the core radio knots, the X-ray jet bends by $\approx$55$^\circ$ \citep{MarscherJ11}.
The spectrum of the kpc-scale jet radiation from the radio to X-ray range cannot be described by the synchrotron radiation spectrum from a single power-law electron energy distribution, therefore, \citet{MarscherJ11} supposed that the X-ray radiation is formed by the inverse Compton scattering of the cosmic microwave background under the assumption of the ultra-relativistic kpc-scale jet directed at a small angle $\left(\approx10^\circ \right)$ to the line of sight \citep[``beamed IC/CMB'' model][]{Tav00, Cel01}.
That is, the same mechanism as it is assumed to act in kpc-scale jets of the core-dominated quasars \citep[see, e.g.,][]{HarrisKraw06}).
For OJ~287, \citet{Meyer19} did not find a contradiction between the ``beamed IC/CMB'' model and the \textit{Fermi}-LAT data, whereas this model predicted a high level of constant flux in the gamma range, which was not detected for quasars 3C~273 \citep{MeyGeor14} and PKS~0637$-$752 \citep{Meyer15}. 
Alternatively, the collection of observed properties of jets of quasars 3C~273 \citep{MBK10}, PKS~1127$-$145 \citep{ButP19}, PKS~0637$-$752, PKS~1045$-$188, and  PKS~1510$-$089 \citep{ButPSN20} was explained in the framework of the model that the X-ray emission is formed due to the inverse Compton scattering of the central source radiation (IC/CS).
In this case, under the radiation of the central source (CS), we mean the pc-scale jet radiation relativistically amplified in the reference frame of the kpc-scale jet.
In this framework, kpc-scale jets are moderately relativistic and form an angle with the line of sight of several tens of degrees.
Assuming that the curvature of the kpc-scale jet OJ~287 is due to the precession, the value of its period is affected by the jet angle with the line of sight.
Thus, by combining data for pc- and kpc-scales, we can conclude about the nature of the central object in OJ~287.

The content of the paper is as follows.
In Section~2, we estimate the physical parameters of the knots, the geometrical and kinematic parameters of the kpc-scale jet OJ~287, assuming that the X-ray radiation of the knots, located before the bend (up to $\approx15^{\prime\prime}$), is formed due to the inverse Compton scattering of the central source radiation.
The analysis is based on \textit{Chandra} observations processed using the latest version of the CIAO~4.13 package and the calibration database CALDB.
In Section~3, we determine the precession period of the central engine based on the curvature of the kpc-scale jet under the assumptions of both IC/CS and ``beamed IC/CMB''.
The discussion of the obtained results and conclusions are in Section~4.

\section{X-ray emission of the kiloparsec-scale jet}
\subsection{\textit{Chandra} observations and processing}

\begin{figure}  
    \centering
    \includegraphics[scale=0.7]{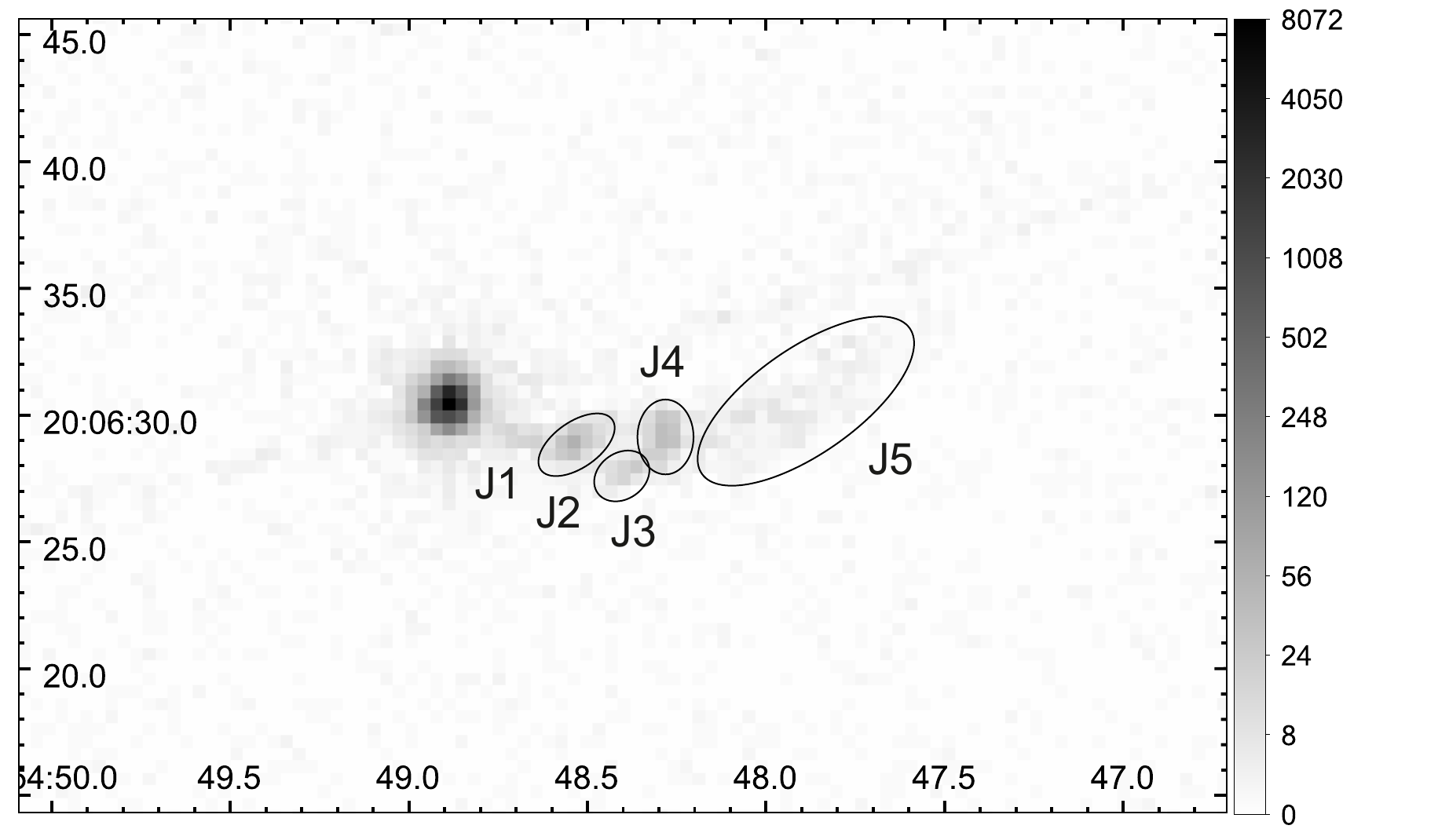}
    \caption{The OJ~287 X-ray jet map in the range of 0.2-6~keV. Knots are denoted according to the nomenclature of \cite{MarscherJ11}. Ellipses mark the regions used to determine the flux and spectral index of radiation. The gray-scale is given in the instrumental units of the photon count rate.}
    \label{fig:imgchandra}
\end{figure}

Using the ACIS-S detector, the \textit{Chandra} X-ray observatory observed OJ~287 once in December 2007, observation number Obs~ID~9182 \citep{MarscherJ11}. 
Since the calibration files for \textit{Chandra} observation processing have been updated, we have re-processed this data using version 4.13 of the CIAO package and version 4.94 of the calibration database CALDB~4.94.
We generated a new event file \textit{evt2} using the standard \textit{chandra\_repro} script.
The resulting file was used in the \textit{deflare} procedure to filter on the light curve noise flares that exceed level 3$\sigma$.
Then, the \textit{dmgti} procedure filtered the data for which the temperature in the detector focal plane did not exceed 156~K.
As a result, the total exposure time of 49.97~kiloseconds (kc) was reduced to 49.77~kc. 
We used the filtered data to calculate the flux from the knots and to model the spectrum in the SERPA package.
Spectrum modeling and flux calculation were performed for the selected in Fig.~\ref{fig:imgchandra} regions that include the knots of the kpc-scale jet OJ~287 except for the first knot J1, which locates close to the bright core, complicating the analysis for it.
We use the knot nomenclature introduced by \citet{MarscherJ11}. We modeled the spectrum in the photon energy range 0.2$-$6~keV under the assumption of a power-law spectrum with a fixed Galaxy absorption.
The neutral hydrogen column density in the blazar direction $N_\text{H}=3.02\cdot10^{20}$~cm$^{-2}$ was calculated using the COLDEN script, based on the data from \citet{DickeyLoc90}. 
The obtained results are shown in Table~\ref{tab:cxcobs}.

\begin{table}
 \caption{
 X-ray observational data for the OJ~287 jet knots.
The columns include: (1) knot; (2) size of a region used in analysis; (3) the distance from the center of the region to the blazar core; (4) integral emission flux in the energy range of 0.2$-$6~keV, in $10^{-14}$~erg~cm$^{-2}$~s$^{-1}$; (5)  spectral flux at the photon frequency, corresponding to the energy of 1~keV, in nJy; (6) X-ray spectral index in the range of 0.2$-$6~keV.
 } 
 \label{tab:cxcobs} 
 \medskip
 \begin{tabular}
 {|c|c|c|c|c|c|}
  \hline
 Knot & Size & $R^{\prime\prime}$ & $F_{0.2-6}$ & $F_1$ & $\alpha_\text{X}$ \\
  \hline
 (1) & (2) & (3) & (4) & (5) & (6)  \\
  \hline
  J2 & ~$1.7\times0.9$~ & ~$5.1$~ & ~$4.69\pm^{0.53}_{0.52}$~ & ~$5.91\pm^{0.58}_{0.50}$~ & ~$0.78\pm0.15$~ \\
  \hline
  J3 & ~$1.2\times0.9$~
   & ~$7.5$~ & ~$5.16\pm^{0.85}_{0.86}$~ & $2.13\pm^{0.35}_{0.36}$ & $0.69\pm0.24$ \\
   \hline
   J4 & $1.1\times1.5$ & $8.4$ & $1.49\pm^{0.13}_{0.15}$ & $6.16\pm^{0.54}_{0.62}$ & $0.79\pm0.14$ \\
   \hline
  J5 & $5\times2$ & $13.2$ & $1.24\pm^{0.12}_{0.14}$ & $5.13\pm^{0.50}_{0.62}$ & $0.78\pm0.16$ \\
  \hline
  \end{tabular}
\end{table}

\subsection{Inverse Compton scattering of the central source emission}

We consider the inverse Compton scattering of the central source emission (IC/CS) using the formulae obtained by \citep{ButP19}.
These formulae determine the flux of scattered radiation under IC of power-law photon spectrum $\left(F\propto\nu^{-\alpha}\right)$ on electrons with power-law energy distribution.

\begin{figure}
    \centering
    \includegraphics[scale=0.65]{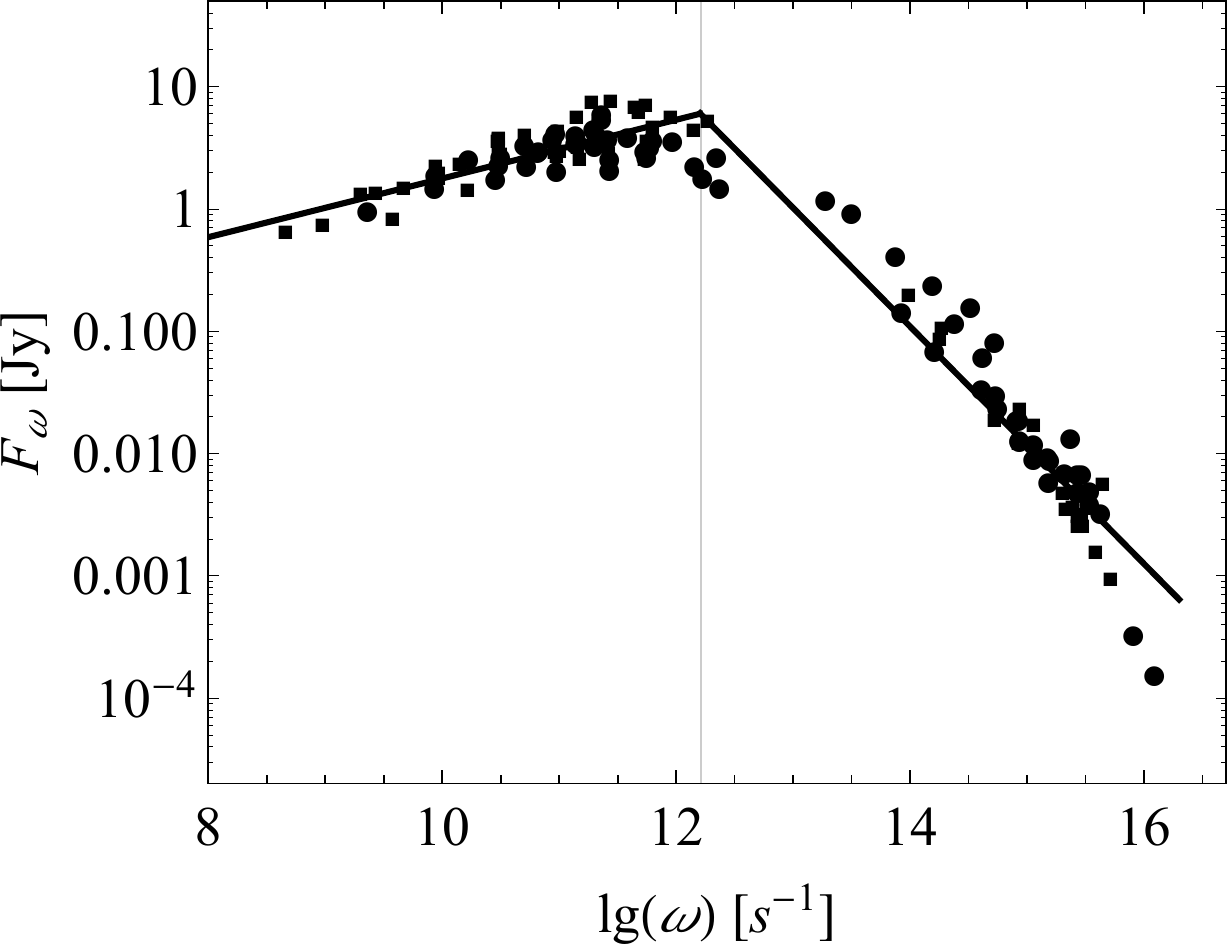}
    \caption{The central source spectrum and its approximation by two power-law parts. The average data at a single frequency is represented by circles and individual observation points are shown by squares. The vertical line marks $\omega_0$.}
    \label{fig:CSspec}
\end{figure}

The observed spectrum of the blazar OJ~287 is more complicated, but that can be approximated by two power parts (Fig.~\ref{fig:CSspec}, Table~\ref{tab:CSsp}). 
Since the central source means the parsec-scale jet radiation, which, at least in the radio-mm range, dominates the radiation of the other parts of the active nucleus \cite{Kovalev05}, the calculation of the CS spectrum in the reference frame of the kpc-scale jet is carried out according to
\begin{equation}
    \omega_\text{j}=\omega\left(1+z\right)\delta_\text{j}/\delta,
    \label{eq:wj}
\end{equation}
where $\omega_\text{j}$ is the photon frequency in the jet's reference frame, corresponding to the photon frequency of  $\omega$ in the observer's reference frame, $z$ is the object redshift, $\delta_\text{j}$ and $\delta$ are Doppler factors for  observers at the Earth and kpc-scale jet.
We estimated the Doppler factor $\delta_\text{j}=\sqrt{1-\beta_\text{pc}^2}/(1-\beta_\text{pc}\cos\theta_\text{pc}^\text{kpc})$ based on the following.
The speed of the jet components is $\beta_\text{pc}=0.9979c$, obtaining from the apparent speed of the fastest moving components \citep{Lister19}. 
Since the CS radiation, scattered within a single knot of the kpc-scale jet, was radiated through thousands of years, the average angle between the pc-scale jet axis and the line of sight, $\theta_\text{pc}=1.9^\circ$, was taken as the angle of the pc-scale jet with the line of sight. 
This value is obtained under the assumption of the helical jet axis precession under the half-opening angle of the precession cone of $0.7^\circ$ and the angle of the precession cone axis with the line of sight of $1.8^\circ$ \citep{ButP20}.
The angle between the pc- and kpc-scale jets, $\theta_\text{pc}^\text{kpc}$, we found from $\theta_\text{pc}$ and the difference of jet position angles on pc- (PA$_\text{pc}$) and kpc- (PA$_\text{kpc}$) scales \citep[see, e.g.,][]{CM93, ButP19, ButPSN20}.
The kpc-scale jet is curved \citep{PerlmanStocke94, MarscherJ11}, so we took for calculations for each considered OJ~287 jet knot of the value of its PA as PA$_\text{kpc}$.
As PA$_\text{pc}$, we took the PA of the helical jet axis, which is equal to $265^\circ$ \cite{ButP20}.
For an accurate estimate of $\theta_\text{pc}^\text{kpc}$, the value of the azimuth angle of the bend is required. 
This angle describes the jet position relative to the plane containing the jet axis and the line of sight, and it can take values from 0 to 360$^\circ$.
Changing the azimuth angle values in the specified interval in increments of 10$^\circ$, we obtained a set of values $\theta_\text{pc}^\text{kpc}$, the median of which was used to calculate $\delta_\text{j}$.
For all knots of the kpc-scale jet $\delta_\text{j}\approx31$. 
According to formulae (3) and (5) in \cite{ButP20}, we found $\delta=19$ as the average value of the Doppler factor of the radiating region in the helical parsec-scale jet.

\begin{table}
 \caption{
Approximations of the CS spectrum by power laws.
 } 
 \label{tab:CSsp} 
 \medskip
 \begin{tabular}
 {|c|c|c|c|}
  \hline
Part & Frequency range, & Coefficient $Q$, & $\alpha$ \\
~ &  s$^{-1}$  & $10^{-23}$~erg~cm$^{-2}$~s$^{-1}$~Hz$^{\alpha-1}$ &\\
  \hline
  1 & $<1.6\cdot10^{12}$ & $\left(7.03\pm^{2.57}_{1.48}\right)\cdot 10^{-3}$ & $-0.24\pm0.02$\\
  \hline
  2 & $>1.6\cdot10^{12}$ & $\left(4.14\pm^{5.75}_{1.52} \right)\cdot10^{12}$ & $0.97\pm0.04$\\
  \hline
  \end{tabular}
\end{table}

Within the framework of IC/CS, the X-ray spectral index $\alpha_\text{X}$ can take two values \citep{ButP19}. 
In the first case, $\alpha_\text{X}=\alpha_i$, where $\alpha_i$ is the spectral index of the $i$-th part of the CS spectrum. 
In this case, the scattered radiation is mainly formed by scattering photons belonging to the power-law spectrum on electrons having an energy corresponding to the energy of one of the boundaries of the power-law electron spectrum (the so-called restriction by the electron spectrum).
In the second case, $\alpha_\text{X}=\alpha_\text{R}=(\gamma-1)/2$, where $\gamma$ is the spectral index of the electron energy distribution.  
In this case, the observed scattered radiation flux is mainly formed due to IC of photons with the frequency corresponding to one of the boundary frequencies of the power-law spectrum on electrons having the energy far from the boundary values (the so-called restriction by the photon spectrum).

For all considered knots of the OJ~287 jet $\alpha_\text{X}=\alpha_\text{R}$ (in the frequency range from 1.4 to 15~GHz) $\alpha_\text{R}=0.8\pm0.1$ \citep{MarscherJ11}. 
The dominant boundary of the photon spectrum in IC is the upper one if $\alpha_\text{R}>\alpha_i$ and the lower one if the inverse inequality holds.
Thus, under IC of both the 1st and 2nd parts of the CS spectrum, IC of photons with a frequency of  $\omega_\text{0, j}=3.3\cdot10^{12}$s$^{-1}$ gives the main contribution to the scattered radiation. 
From the formula (6) in \citep{ButP19} with the substitution $\omega_\text{0, j}$ and the parameters of the 1st part of the OJ~287 CS spectrum (since at these frequencies there is less contribution from other parts of the active nucleus), we estimated the average magnetic field strength $B=1.1 \cdot10^{-7}$~G and the electron number density $n_e=25\Gamma_\text{min}^{-1.6}$~cm$^{-3}$ ($\Gamma_\text{min}$ is the minimum electron Lorentz factor) for knots J2, J3, and J4. 
For this, we used the observational data from Tables~1 and \citep{MarscherJ11}, $\gamma=2.6$, the kpc-scale jet angle with the line of sight. $\theta_\text{kpc}=35^\circ$ (see Section~2.3), and $\Lambda$CDM-model with parameters $\Omega_\text{m}=0.27$, $\Omega_\Lambda=0.73$, $H_0=71$~km~s$^{-1}$~Mpc$^{-1}$ \citep{Komatsu09}. 
It is seen that the values $n_e$ and $B$ are acceptable for this type of source. 

\subsection{The viewing angle and speed of kiloparsec-scale jet}

X-ray radiation within the extended region J5 has a low intensity without a pronounced spatial distribution.
Therefore, we assume that in this region, the X-ray radiation is formed due to IC of the cosmic microwave background photons. 
Meanwhile, in other nearer to CS knots, X-ray radiation is generated by IC/CS (see Section.~2.2). 
For knot J4, the domination of the X-ray flux formed due to IC/CS over the flux produced due to IC/CMB allows us to determine the lower limit on the viewing angle of the jet part, containing knot J4, with the line of sight \citep[see formula (13) in][]{ButP19}
\begin{equation}
    \theta_\text{kpc}\geq\left[\frac{2^{\gamma+1}|2\alpha_1+1-\gamma|}{\gamma+3} W_\text{CMB}\frac{4\pi c R_\text{J4}^2}{L_\text{CS}} \left( \frac{\omega_\text{CMB}}{\omega_\text{0, j}}\right)^{(\gamma-1)/2-1}\right]^{1/(\gamma+3)}\geq35.2^\circ,
    \label{eq:thkpc}
\end{equation}
where $W_\text{CMB}=1.2\cdot 10^{-12}$~erg~cm$^{-3}$ and $\omega_\text{CMB}$ are the energy density and frequency of the cosmic microwave background maximum at the object redshift $z$, $R_\text{J4}$ is the knot J4 distance from the central source, $L_\text{CS}=4\pi (1+z)^{3+\alpha_1}(\delta_\text{j}/\delta)D^2_L Q_1 \omega_\text{0, j}^{-\alpha_1+1}\approx5.2\cdot 10^{47}$~erg~s$^{-1}$ is the CS luminosity in the reference frame of the kpc-scale jet ($D_L=1576.8$~Mpc).
Using $R_\text{J5}$ instead of $R_\text{J4}$ in the inequality inverse to (\ref{eq:thkpc}), we obtained an upper limit on the kpc-scale jet angle with the line of sight $\theta_\text{kpc}\leq42^\circ$.

Matching of pc- ($\theta_\text{pc}=1.9^\circ$) and kpc-scale ($\theta_\text{pc}=1.9^\circ$) jet angles with the line of sight is possible by relativistic aberration (see Section~2.2).
Namely, when the jet decelerates, the angle between the jet velocity vector and the line of sight in the observer's reference frame increases.
Then the Doppler factor of the kpc-scale is \citep{ButP19}
\begin{equation}
    \delta_\text{kpc}=\delta_\text{pc}\frac{\sin \theta_\text{pc}}{\sin \theta_\text{kpc}}=1.02.
    \label{eq:relab}
\end{equation}
Then the speed of the kpc-scale jet can be equal to 0.025$c$ or 0.971$c$. 
Since the appearances of jet deceleration are already detected at distances of 100~pc from the core \citep{Homan15}, the first value seems more likely.

\subsection{Expected gamma-ray flux from the jet}

\begin{figure} 
    \centering
    \includegraphics[scale=0.5]{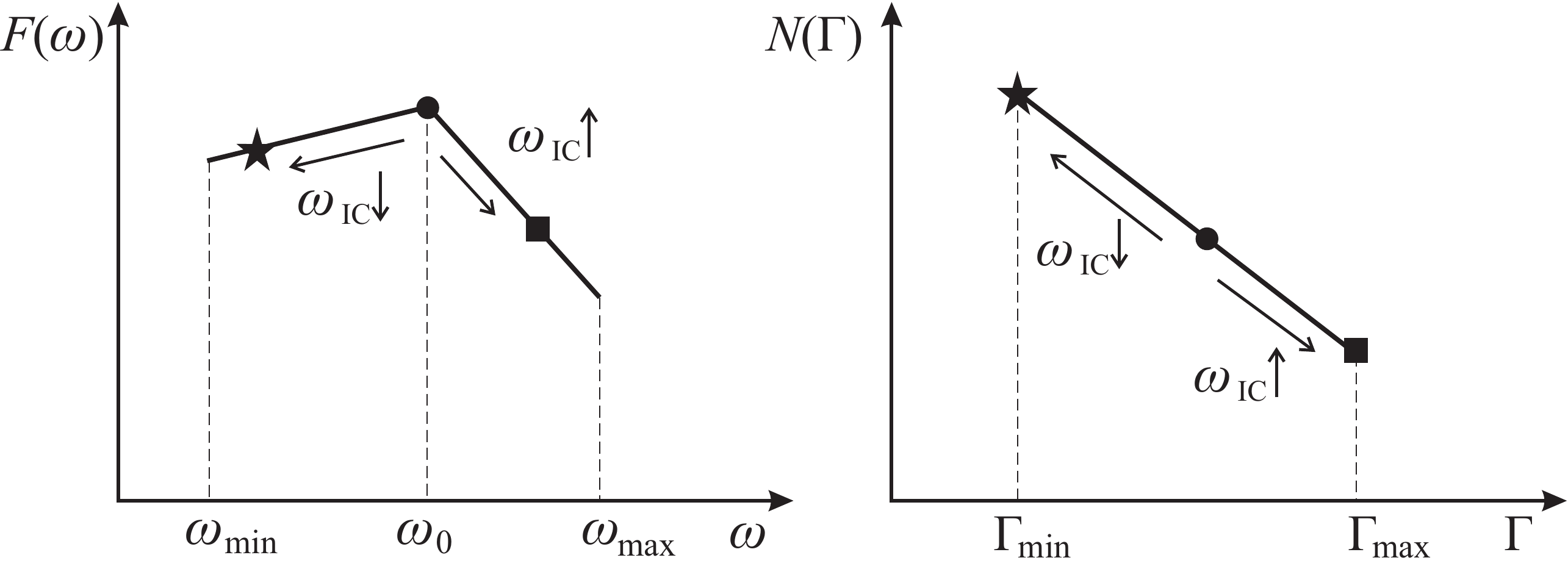}
    \caption{Schemes of the photon and electron spectra. The same symbols indicate the spectral parts, which contribute mainly to the scattered radiation flux at different frequencies. The filled circle corresponds to the energy of interacting particles that produce radiation in the photon energy range 0.2$-$6~keV, the square corresponds to the energies of the particles radiated at higher frequencies, the asterisk --- at lower frequencies.}
    \label{fig:scheme_e_ph}
\end{figure}

\begin{figure}  
    \centering
    \includegraphics[scale=0.8]{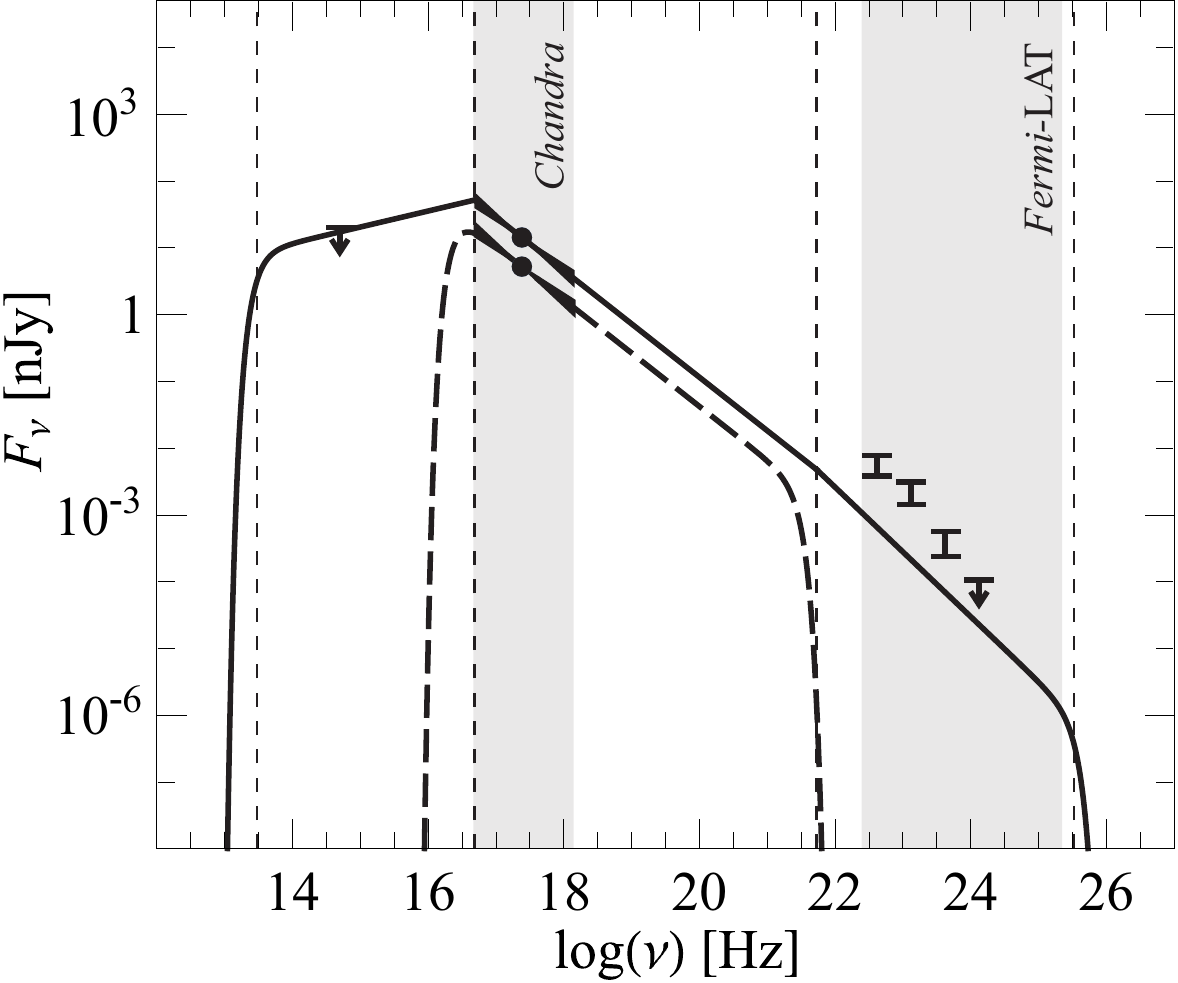}
    \caption{The simulated spectrum of high-frequency radiation of the OJ~287 jet knots. The total radiation from knots J2, J3, and J4 under IC/CS is shown by a solid line, the radiation from knot J5 under IC on the cosmic microwave background for the considered sub-relativistic jet is displayed by a dashed line. The points and triangles show the observed X-ray fluxes at the photon energy of 1~keV and the measurement errors of the X-ray spectral index, respectively. The dotted vertical lines indicate the break frequencies in the knot spectrum under IC/CS. The shaded regions mark the operating ranges of \textit{Chandra} and \textit{Fermi}-LAT. The short black lines show the fluxes and the upper limit on them in the optical and gamma ranges. The electron energy spectrum was assumed to be power-law in the range from $\Gamma_\text{min}=300$ to $\Gamma_\text{max}=10^5$.}
    \label{fig:gammaspec}
\end{figure}

The flux in the analyzed frequency range, corresponding to the photon energies of 0.2$-$6~keV, is formed by IC of photons with a frequency of $\omega_\text{0, j}$.
The energies of the interacting particles are schematically marked with filled circles in Fig.~\ref{fig:scheme_e_ph}. 
Since the scattered photon frequency is uniquely determined by the photon frequency before scattering $\omega_\text{j}$ and the electron Lorentz factor $\Gamma$ \citep[e.g.,][]{Pachol} 
\begin{equation}
\omega_\text{IC}=k_\text{IC} \omega_\text{j} \Gamma^2,
    \label{eq:wic}
\end{equation}
where $k_\text{IC}=4/3/(1+z)$, IC of photons with the frequency of $\omega_\text{0, j}$ acts in the interval from $\omega_\text{X, br1}=k_\text{IC} \omega_\text{0, j}\Gamma_\text{min}^2$ to $\omega_\text{X, br2}=k_\text{IC} \omega_\text{0, j}\Gamma_\text{max}^2$, where $\Gamma_\text{min}$ and $\Gamma_\text{max}$ are the Lorentz factors of the lower and upper bounds of the power-law electron energy spectrum, respectively. 
At frequencies higher than $\omega_\text{X, br2}$, the scattered radiation is formed by IC of photons of part 2 of the CS spectrum on electrons with $\Gamma_\text{max}$ and has a spectral index equal to $\alpha_2$.
In Figure~\ref{fig:scheme_e_ph}, the energies of the interacting particles for this case are marked with filled squares.
For the constant energy of scattering electrons, the growth of $\omega_\text{IC}$ is provided by the scattering of photons with an increasing frequency.
Scattering of CS photons with the maximum frequency of $\omega_\text{max, j} $ produces scattered radiation at the frequency of $\omega_\text{X, max}=k_\text{IC}\omega_\text{max, j}\Gamma_\text{max}^2$, above which the high-energy radiation spectrum cuts off.
At frequencies below $\omega_\text{X, br1}$, the scattered flux is formed by IC of photons of the part~1 of the CS spectrum on electrons with $\Gamma_\text{min}$ and has a spectral index equal to $\alpha_1$.
In Figure~\ref{fig:scheme_e_ph}, stars denote energies of the interacting particles. 
Figure~\ref{fig:gammaspec} shows the simulated spectrum of the scattered radiation under IC/CS.
The upper limits on the flux from the kpc-scale jet in the optical \citep{MarscherJ11} and gamma \citep{Meyer19} ranges are given too.
The spectrum is plotted with the parameters: $\Gamma_\text{min}=300$ and $\Gamma_\text{max}=10^5$, the latter of which was chosen so that, for a magnetic field of $\sim10^{-6}$~G, there was no synchrotron radiation at a frequency above 15~GHz.
The value of $\Gamma_\text{max}$ defines the frequencies $\omega_\text{X, br2}$ and $\omega_\text{X, max}$.
Figure~\ref{fig:gammaspec} shows that a change in these frequencies will not lead to an excess of the model flux of scattered radiation in the gamma range over the values set as the upper limit on the observed flux.
The requirement for the absence of a break in the spectrum of scattered radiation at frequencies corresponding to the energy 0.2$-$6~keV gives $\Gamma_\text{min}\leq300$.
On the other hand, $\Gamma_\text{min}\gtrsim300$ is necessary for agreement with the upper limit on the optical flux.
Note that the inverse spectrum at low radio frequencies is explained by the absorption acting in the pc-scale jet medium \cite{Slish63, Lobanov98, Pushkarev12}. 
The fact that the OJ~287 low-frequency spectrum is not as steep as expected under absorption may be explained by the fact that,  at low frequencies, the angular resolution of a single-dish antenna is not sufficient to separate the radiation from a compact core and a kiloparsec-scale jet.
Therefore, the kiloparsec-scale jet radiation significantly contributes to the observed total radiation of the blazar OJ~287 at frequencies belonging to part 1 of the CS spectrum.
Then the true CS spectrum at these frequencies can be steeper, and therefore the scattered radiation spectrum at frequencies from $\omega_\text{X, min}$ to $\omega_\text{X, br1}$ is initially steeper.
Then correspondence to the upper limit on the optical flux will be held at lower $\Gamma_\text{min}$.

\section{Precession of the kpc-scale jet}
\label{sec:prec}
For the interpretation of the properties of the long-term optical variability and the VLBI observations of the OJ~287 jet, the precession of the central engine and the helical jet are often assumed.
Therefore, it is natural to expect the detection of a sign of precession on the kpc-scales, which would reveal itself as a curved jet.
A similar curved jet is observed as far as 16$^{\prime\prime}$ from the core (X-ray radiation is detected from this part, see Fig.~\ref{fig:imgchandra}), and the bend continues to 27.4$^{\prime\prime}$ where the farthest feature of the kpc-scale jet is detected at 1.4~GHz frequency observations \citep{PerlmanStocke94}.
Assuming that the maximum difference in PA observed for this knot and J1 is twice the deviation PA of the precession cone axis ($2\Delta \text{PA}=53^\circ$), the ratio of the angle of the precession cone axis with the line of sight $\theta_\text{0, kpc}$ to the precession cone half-opening angle $\xi_\text{kpc}$ can be found from the expression (3) in \citep{But18a}:
\begin{equation}
   \text{tg}\Delta \text{PA}=\left[\left(\frac{\theta_\text{0, kpc}}{\xi_\text{kpc}} \right)^2-1 \right]^{-1/2}. 
    \label{eq:tgDPA}
\end{equation}
We obtained $\theta_\text{0, kpc}/\xi_\text{kpc}=2.25$, which roughly corresponds to a similar ratio $\theta_\text{0, pc}/\xi_\text{pc}=2.57$ for pc-scales ($\theta_\text{0, pc}=1.8^\circ$, $\xi_\text{pc}=0.7^\circ$ \cite{ButP20}).
Using the formula~(\ref{eq:relab}) for $\theta_\text{pc}=\theta_\text{0, pc}+\xi_\text{pc}$, we found the half-opening angle of the kpc-scale jet precession cone $\xi_\text{kpc}\approx 18^\circ$.

The distance $\Delta R$ between the farthest knot in the radio band and the knot J1 is the half-wavelength of the precession-curved jet in the projection on the plane of the sky.
The analysis of X-ray observations gives the necessary angle with the line of sight $\theta_\text{kpc}$ and the speed $\beta_\text{kpc}$ of the kpc-scale jet.
Then the precession period in the observer's reference frame is 
\begin{equation}
    T_\text{kpc}=\frac{2\Delta R}{\beta_\text{kpc}c \sin \theta_\text{kpc}}.
    \label{eq:Tkpc}
\end{equation}
To calculate $T_\text{kpc}$ using the formula (\ref{eq:Tkpc}), the following values are required $\theta_\text{kpc}$ and $\beta_\text{kpc}$. 
The estimation of these parameters can be obtained from the analysis of the X-ray formation mechanisms of the kpc-scale jet.
At the moment, there are two rival mechanisms, giving fundamentally different estimates of $\theta_\text{kpc}$ and $\beta_\text{kpc}$.
In the framework of IC/CS, $\theta_\text{kpc}\approx38^\circ$ and $\beta_\text{kpc}=0.025$. 
On the other hand, the ``beamed IC/CMB'' model implies an ultra-relativistic kpc-scale jet. 
For it, according to one estimate $\delta_\text{kpc}=22.5$ and $\theta_\text{kpc}=2.6^\circ$~\cite{Meyer19}, which is impossible for real $\beta_\text{kpc}$, by the other estimate $\delta_\text{kpc}=8$ and $\theta_\text{kpc}=3.8^\circ$ before the bend and 7$^\circ$ for the X-ray jet end \cite{MarscherJ11}. 
The last parameter set assumes $\beta_\text{kpc}=0.97$, which we use in further calculations. 
Thus, for IC/CS $T_\text{kpc}^\text{IC/CS}=4.5\cdot10^7$~years, for ``beamed IC/CMB'' $T_\text{kpc}^\text{IC/CMB}=8.1\cdot10^6$~year.

\begin{figure}  
    \centering
    \includegraphics[scale=0.5]{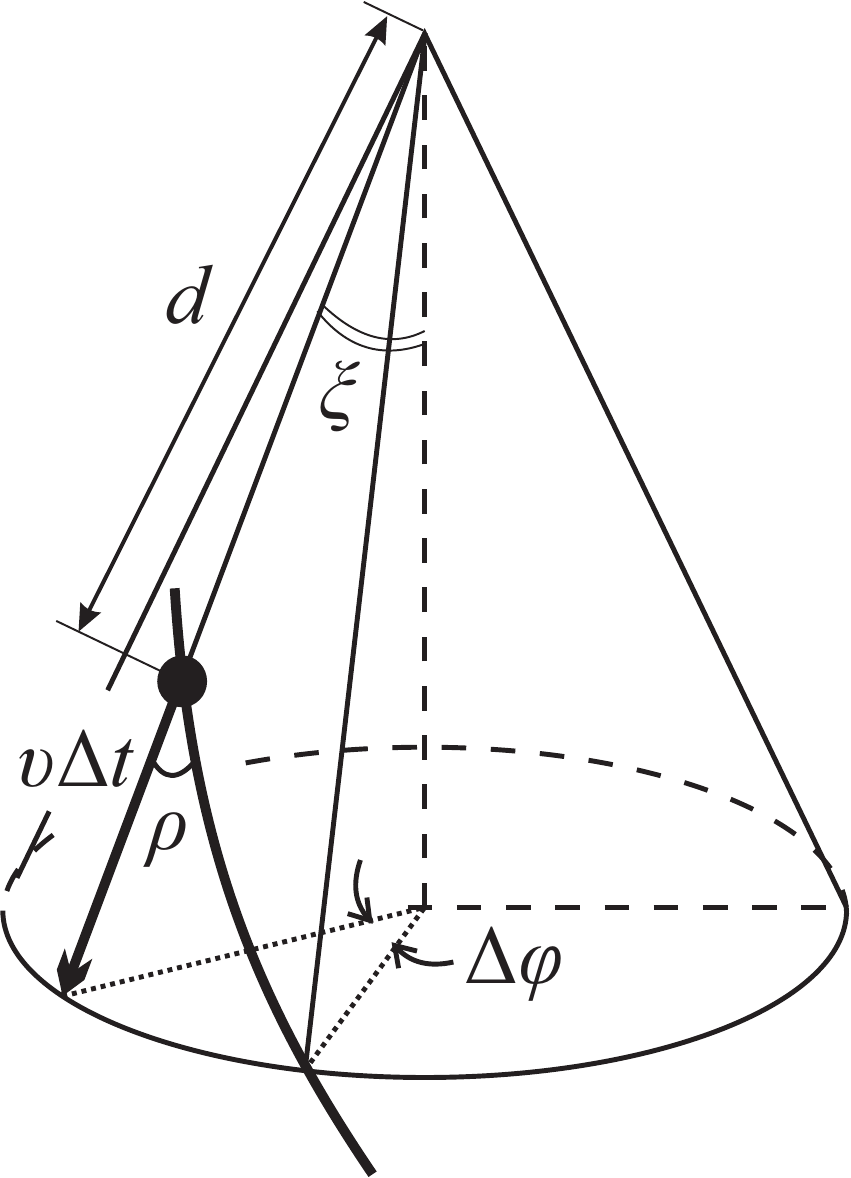}
    \caption{Scheme illustrating the change in the azimuth angle at a fixed distance from the cone apex when the jet components move along the precession cone generatrix. It is shown one jet component (the filled circle) and a jet part near it (the thick line). Geometrical and kinematic parameters, using for estimation of $\Delta\varphi$, are denoted.}
    \label{fig:coneT}
\end{figure}

The precession period, determined by optical data of 12 years, does not appear in the light curves in the radio range.
Moreover, the periods of the radio flux variability and change in the inner jet PA approximately correspond to both each other and $\approx25-30$~years \citep{Britzen18,Ryabov16,Sukharev19},
and allow us to correlate their formation with the jet helical shape due to precession. 
These periods are significantly smaller than the periods found here. But taking into account the different speeds and scales of the jets, it may be possible to coincide them.
To do this, we find an expression for the period in terms of the azimuth angle change at some distance from the precession cone apex (see Fig.~\ref{fig:coneT}). 
The region radiating at a given frequency is located at a constant distance $d$ from the cone apex. Note that the cone apex may not coincide with either the jet base or the BH position. Then the small change in the azimuth angle at the distance of $d$ is
\begin{equation}
    \Delta\varphi\approx\frac{\upsilon\Delta t \, \tan \rho}{\left( d+\upsilon\Delta t\right)\cos \xi},
    \label{eq:dfi}
\end{equation}
where $\rho$ is the the swirl angle of the jet matter. 
Given that $d \gg \upsilon \Delta t$ and $T=n \Delta t$, we find the period ratio of the quantities corresponding to the pc- and kpc-scales
\begin{equation}
    \frac{T_\text{pc}}{T_\text{kpc}}=\frac{d_\text{pc}\, \beta_\text{kpc} \cos \xi_\text{pc}}{d_\text{kpc}\, \beta_\text{pc} \cos \xi_\text{kpc}}.
    \label{eq:ratioT}
\end{equation}
The authors of \cite{ButP20} explained the difference between the periods in the optical and radio ranges by the non-radial motion of the jet components and that the region responsible for optical radiation is closer to the true jet base than the VLBI core and the pc-scale jet. 
As can be seen from the formula~(\ref{eq:ratioT}), the discrepancy between the periods in the optical and radio ranges under radial motion can also be explained by the different distances from the cone apex to the regions responsible for the observed values. 
The only parameter in the expression~(\ref{eq:ratioT}), which has no reliable estimate, is $d_\text{pc}$.
For the VLBI core, it can be estimated at $d<10.3$~pc~\citep{ButP20} out of the astrometric shift of the VLBI core position at different observation frequencies~\citep{Pushkarev12}.
Then for the optically emitting region, we can assume $d_\text{pc}\approx4$.

\begin{figure} 
    \centering
    \includegraphics[scale=0.9]{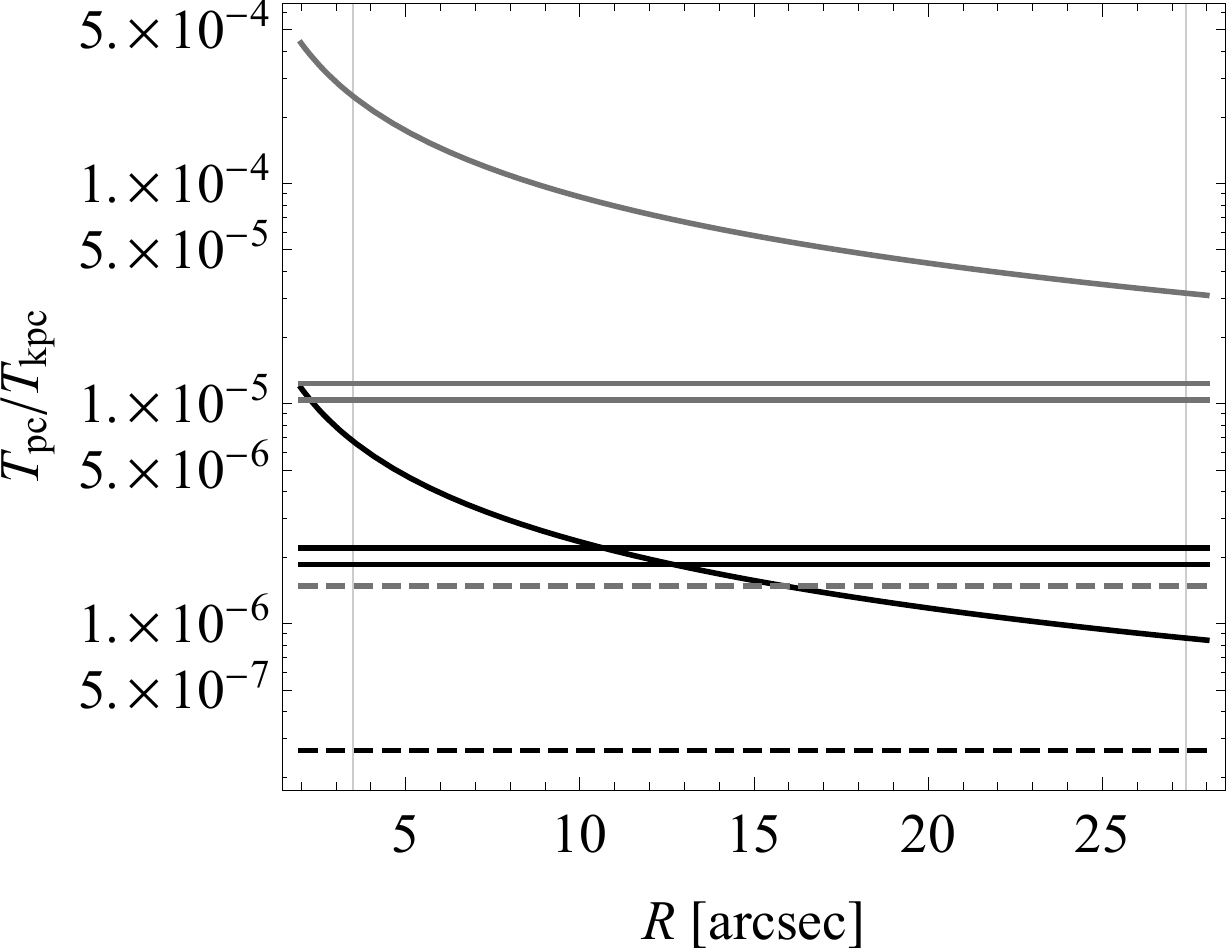}
    \caption{The ratio of defined at pc- and kpc-scales precession periods. Black color corresponds to values, which obtained under IC/CS, gray color --- to values for ``beamed IC/CMB’’. The curved lines show expression~(\ref{eq:ratioT}) as a function of the angular distance from the core with the substitutions:  $\xi_\text{pc}=0.7^\circ$, $\xi_\text{kpc}=18.1^\circ$, $\beta_\text{pc}=0.9979$, $\beta_\text{kpc}=0.025$ (for IC/CS) and $\xi_\text{pc}=0.7^\circ$, $\xi_\text{kpc}=2.3^\circ$, $\beta_\text{pc}=0.9979$, $\beta_\text{kpc}=0.97$ (for ``beamed IC/CMB’’). The straight lines mark the corresponding ratios of the periods $92\pm8$ (solid) and 12~years (dashed) to the values $T_\text{kpc}$ found in the Section~\ref{sec:prec}.}
    \label{fig:TpcKpc}
\end{figure}

On the one hand, the OJ~287 precession period of 12 years in the observer's reference frame is considered over more than 30 years. 
On the other hand, this periodicity can occur in a jet having the helical shape due to the development of hydrodynamic instabilities, while the true precession period is reflected in the modulation of the flux peak values during 12-year flares~\citep{ButP20}. 
Under this assumption, the precession period of the OJ~287 central engine was estimated at $ 92 \pm8$~years in the observer's reference frame, which corresponds to 1200~years in the source reference frame~\cite{ButP20}. 
To check whether the periods related to the pc-scales values correspond to the period found by the kpc-scale jet curvature, we expressed the value of $d_\text{kpc}$ through the angular distance $R$, which was changed from 3 to 28$^{\prime\prime}$, i.e., to the distance at which the OJ~287 kpc-scale jet stops being detected in the radio range.
The ratio of pc- and kpc-scale periods expected from formula~(\ref{eq:ratioT}) for IC/CS and the ``beamed IC/CMB'' model is shown in Figure~\ref{fig:TpcKpc}.
It is seen that under the IC/CS assumption, the precession period determined by the kpc-scale jet is consistent with the precession period of the helical jet of 92~years~\citep{ButP20}.

\section{Discussion and conclusions}
\label{sec:dc}

OJ~287 is the first object for which the existence of a binary black hole system was assumed based on the periodicity of the long-term optical light curve \citep{Sillanpaa88}. The OJ~287 non-thermal spectrum with weak radiation in the lines implies that almost all observed optical radiation is formed in the relativistic jet. Then the variability period can naturally be explained by the jet viewing angle change due to the jet helical shape, which is caused by the central engine precession \citep{Katz97,VR98, ValtonenPihajoki13}. But anyway, to obtain such a small precession period, it is necessary to assume a binary BH at the center of OJ~287.

Long-term VLBI observations of the OJ~287 pc-scale jet, performed using \textit{Very Long Baseline Array} (USA) at frequencies 43~GHz \citep{Agudo12} and 15~GHz~\citep{Britzen18,ButP20}, detect the jet position change with a (quasi) period of $\approx22-28$~years. The $\approx25$ year period of long-term radio flux variability is in good agreement with this value. The difference in the periods determined from the optical and radio data is explained by the fact that the radiation of the corresponding frequencies comes from different parts of the jet: the region emitting in the optical range is closer to the true jet base than the VLBI core \citep{ButP20}.

\citet{Britzen18} showed that the origin of the period of $\approx25$~years is possible both in the system of binary BH and, under some parameters, in the single BH system. On the other hand, \citet{ButP20} showed that the periods observed in the radio and optical range can be formed in a helical jet, which acquired this form due to the development of the Kelvin-Helmholtz instability. The precession of this helical jet can be reflected in the difference in peak fluxes of 12-year flares. The precession period found under these assumptions was 92~years in the observer's reference frame, which corresponds to 1200~years in the source's reference frame. Precession with this period can occur in the system of a single supermassive BH and its accretion disk \citep{ButP20}. It is impossible to give an advantage to one of the two assumptions about the OJ~287 central engine based on the available photometric optical and VLBI data. On the other hand, the blazar OJ~287 has a $\approx28^{\prime\prime}$ curved kpc-scale jet. Therefore, assuming that this curvature is due to the central engine precession, we can find the period of the kpc-scale jet and, in agreement with the values related to the pc-scale, conclude that the single or binary BH is in the OJ~287 center.
It is important to note that recent studies of the X-ray spectrum of several quasars have found no evidence of binary supermassive BH presence \cite{Saade20}.

For this aim, it is necessary to know the kpc-scale jet speed and angle with the line of sight. These parameters can be estimated from the analysis of kpc-scale jet X-ray emission. But at the moment, there are two assumptions about the formation mechanisms of X-ray radiation for the OJ~287 jet. These are IC/CS and ``beamed IC/CMB'' \citep{Tav00,Cel01}. The latter has been widely used to interpret the properties of kpc-scale jets of core-dominated quasar, but in the light of the data obtained by \textit{Fermi}-LAT, it has already been disproved for several objects \citep{MeyGeor14,Meyer15}, excepting OJ~287 \citep{Meyer19}. Therefore, for the analysis, we also used data on the speed and orientation of the kpc-scale jet, obtained in the framework of the ``beamed IC/CMB''. IC/CS was considered for several objects and gave an interpretation of the observed brightness distributions along the jet, the similarity and difference of the spectral indices determined in the radio and X-ray ranges, without additional assumptions \citep{ButP19,ButPSN20}. 

The main obtained results are following.:

(i) It is shown that IC/CS is a possible X-ray emission mechanism for the OJ~287 kpc-scale jet. Under this, the expected gamma-ray flux does not exceed the upper limit on its set by \textit{Fermi}-LAT data \citep{Meyer19}.

(ii) In the framework of IC/CS, the speed and angle of the kpc-scale jet with the line of sight are 0.025$c$ and $\approx38^\circ$, respectively.

(iii) The accordance of the periods found on the basis of pc- and kpc-scale data is present only for the period obtained within the framework of IC/CS. Under this, the period of the kpc-scale jet helix agrees well only with the precession period of 92~years (in the observer's reference frame), expected under the assumption of a single supermassive BH at the center of OJ~287.

\begin{acknowledgments}
The research was fully supported by Russian Science Foundation, project No. 19-72-00105. 
\end{acknowledgments}

\bibliography{OJ287Xjet}{}
\bibliographystyle{aasjournal}

\end{document}